\documentclass[a4paper,12pt]{article}

\usepackage[utf8]{inputenc}
\usepackage[T1]{fontenc}
\usepackage[margin=1in]{geometry}

\usepackage{textcomp}
\usepackage{lmodern}
\usepackage{ae}
\usepackage[english]{babel}
\usepackage[intlimits]{amsmath}
\usepackage{amssymb}
\usepackage{bm}
\usepackage{amsthm}
\usepackage{floatrow}
\usepackage{exscale}
\usepackage{graphicx}
\usepackage{titlesec}
\usepackage{algorithm}
\usepackage{algorithmic}
\usepackage{listings}
\usepackage{mathtools}
\usepackage{hyperref}
\usepackage{cleveref}
\usepackage{array}
\usepackage{booktabs}
\usepackage{siunitx}
\usepackage{etoolbox}
\usepackage{floatrow}
\usepackage{subfig}

\makeatletter
\providecommand{\@LN}[2]{}
\makeatother

\newfloatcommand{capbtabbox}{table}[][\FBwidth]
\newcommand{\Mod}[1]{\ (\mathrm{mod}\ #1)}

\DeclarePairedDelimiter\abs{\lvert}{\rvert}%
\DeclarePairedDelimiter\norm{\lVert}{\rVert}%
\DeclareMathOperator*{\argmin}{arg\,min}
\makeatletter
\let\oldabs\abs
\def\abs{\@ifstar{\oldabs}{\oldabs*}}
\let\oldnorm\norm
\def\norm{\@ifstar{\oldnorm}{\oldnorm*}}
\makeatother

\usepackage{csquotes}

\usepackage[style=phys,citestyle=phys,giveninits=true,maxbibnames=6,uniquename=init,minbibnames=1]{biblatex}
\newcommand{\uest}{\widehat{u}}
\newcommand{\uref}{u^{\mathrm{ref}}}
\newcommand{\ahat}{\widehat{\alpha}}
\newcommand{\bhat}{\widehat{\beta}}

\newcommand{\tvs}{{\rm TV}_{\rm S}}
\newcommand{\hattvs}{\hat{S}_{{\rm S}}}
\newcommand{\tvt}{{\rm TV}_{\rm T}}
\newcommand{\hattvt}{\hat{S}_{{\rm T}}}

\begin{document}
	
	\title{Data-driven regularization parameter selection in dynamic MRI}
	
	\author{Matti Hanhela$^1$\thanks{matti.hanhela@uef.fi} \and 
		Olli Gr\"ohn$^2$ \and
		Mikko Kettunen$^2$ \and
		Kati Niinim\"aki$^3$ \and
		Marko Vauhkonen$^1$ \and
		Ville Kolehmainen$^1$}
	\date{$^1$ Department of Applied Physics, University of Eastern Finland, 70211 Kuopio, Finland\\
		$^2$ A.I. Virtanen Institute for Molecular Sciences, University of Eastern Finland, 70211 Kuopio, Finland\\
		$^3$ Xray Division, Planmeca Oy, Asentajankatu 6, 00880 Helsinki, Finland\\
		\today\\
	}
	
	\maketitle
	
	\begin{abstract}
		In dynamic MRI, sufficient temporal resolution can often only be obtained using imaging protocols which produce undersampled data for each image in the time series. This has led to the popularity of compressed sensing (CS) based reconstructions. One problem in CS approaches is determining the regularization parameters, which control the balance between data fidelity and regularization. We propose a data-driven approach for the total variation regularization parameter selection, where reconstructions yield expected sparsity levels in the regularization domains. The expected sparsity levels are obtained from the measurement data for temporal regularization and from a reference image for spatial regularization. Two formulations are proposed. Simultaneous search for a parameter pair yielding expected sparsity in both domains, and a sequential parameter selection using the S-curve method. The approaches are evaluated using simulated and experimental DCE-MRI. In the simulated test case, both methods produce a parameter pair and reconstruction that is close to the RMSE optimal pair and reconstruction. In the experimental test case, the methods produce almost equal parameter selection, and the reconstructions are of high perceived quality. Both methods lead to a highly feasible selection of the regularization parameters in both test cases while the sequential method is computationally more efficient.
	\end{abstract}
		
	%%%%%%%%%%%%%%%%%%%%%%%%%%%%%%%%%%%%%%%%%%%%%%%%%%%%%%%%%%%%%%%%%
	\newpage
	\section{Introduction.}
	\label{sec:introduction}
	Dynamic magnetic resonance imaging is used to study hemodynamics, microvascular structure and function by contrast agent or stimulus related changes in a time series of MR images. High spatial and temporal resolution of the dynamic image series is often required for accurate analysis {of the contrast agent or stimulus dynamics.} In many cases, sufficient time resolution can only be obtained by utilizing an imaging protocol which produces undersampled data for each image in the time series. This, however, has the complication that reconstructing undersampled datasets with conventional MR image reconstruction methods, such as regridding \cite{Jac+91}, lead to noisy image series with poor spatial resolution.
	
	Recently, the compressed sensing (CS) framework has led to significant advances in MRI with undersampled data. The theory of CS states that a target signal or image that is sparse in some basis, which is also incoherent with the measurement basis, can be perfectly reconstructed from undersampled data with a high probability \cite{CRT06,Don06,LDP07}. Compressed sensing based approaches have been developed for numerous applications in both static and dynamic MRI, see for example the review \cite{JFL15}.
	
	Provided that the temporal resolution of the MR image series is high enough, one can expect high redundancy in the image series in the sense that the image intensity changes between successive image frames are small and occur only in some parts of the image domain. Therefore a compressed sensing approach based on the sparsity of the time derivative of the images is warranted.
	
	The basic structure in the CS approaches to dynamic MRI is to reconstruct the whole time series of images simultaneously using an appropriate joint reconstruction formulation where a temporal regularization functional is employed for coupling the undersampled data across the time series of images. The most popular approach has been to use total variation (TV) regularization to promote sparsity of the spatial and temporal derivatives of the images, leading to a formulation \cite{AV94,Adl+09}
	\begin{equation}
	\label{eq:regularization_general}
	\uest = \argmin_u \left\lbrace D(u, m) + \alpha\,  \tvs (u) + \beta\, \tvt (u) \right\rbrace
	\end{equation}
	where $u=\{u^1,u^2, \ldots ,u^{T} \}$ denotes the time series of unknown images, $m =\{m^1,m^2, \ldots ,m^{T} \}$ is the MRI data time series, $T$ is the number of time frames, $D$ is the data fidelity term, $\tvs$ is the spatial total variation regularization functional, $\tvt$ is the temporal TV regularization functional and $\alpha$ and $\beta$ are the spatial and temporal regularization parameters, respectively.
	
	The selection of the regularization parameters is crucial in terms of resulting image quality. Classical parameter selection methods, such as the Morozov discrepancy principle \cite{Mor68}, exist, but these have been designed for the selection of a single regularization parameter and derived for a restricted class of problems. Alternatively, the parameter selection can be done by the L-curve method \cite{HO93},  Monte-Carlo SURE \cite{RBU08,Ram+13,Wel+14} or by subjective visual assessment of the reconstructions.
	
	In this work, we propose a data-driven approach for the selection of the regularization parameters $\alpha$ and $\beta$ in \eqref{eq:regularization_general}. The proposed idea is to select a parameter pair which produces an \textit{a priori} expected level of sparsity in both the domain of the spatial TV regularization and the temporal TV regularization.
	
	As a part of the dynamic MRI measurement protocol, fully sampled cartesian measurements of the target are typically taken before the dynamic measurements. The anatomical MRI reconstruction is often used only for visualization, but it can also be used to obtain an \textit{a priori} estimate for the spatial sparsity of the target. In case an anatomical reference image is not acquired, the expected spatial sparsity could also be estimated from a static reconstruction of a long sequence of baseline of the dynamic data or from an anatomical atlas image.
	
	An \textit{a priori} estimate for the expected sparsity in the temporal regularization domain can be extracted from the dynamic MRI data. More specifically, the DC component of the k-space data gives the information about the integral intensity of the unknown image at each frame. This information can then be used to approximate the expected sparsity level of the time derivative of the images when the image changes are based on contrast changes instead of tissue movement. The approach leads to a two-dimensional search from a set of reconstructions over a grid of parameter values $(\alpha,\beta)$.
	
	The idea in the proposed method is a 2D extension of the S-curve method, which was originally proposed for the selection of regularization parameter in sparsity promoting regularization in \cite{Kol+12} and later applied to parameter selection in static x-ray tomography in \cite{Ham+13,Nii13,Nii+16}. The S-curve method was also applied to spatial TV regularization parameter selection in DCE-MRI in \cite{Nii+19}.
	
	Since the proposed 2D parameter search can require a large number of reconstructions over a grid of values of $(\alpha,\beta)$, we also consider a second method where the parameters $\alpha$ and $\beta$ are selected separately by applying the S-curve method sequentially to first the temporal regularization parameter $\beta$ and then the spatial regularization parameter $\alpha$. 
	
	The proposed parameter selection approaches are evaluated using simulated and experimental DCE-MRI data from a rat brain study. We compare the proposed approaches to the L-curve \cite{HO93} and Monte-Carlo SURE \cite{RBU08,Ram+13,Wel+14} parameter selection methods and in the simulated case, also to the true target.
	
	In DCE-MRI, a bolus of gadolinium based contrast agent is injected into the blood circulation and a time series of MRI data with an appropriate $T_1$-weighting is measured to obtain a time series of 2D (or 3D) images which exhibit contrast changes induced by concentration changes of the contrast agent in the tissues. DCE-MRI is used for example in treatment monitoring of breast cancer \cite{Mar+04,Pic+05} and glioma \cite{Pil+15}, and to detect perfusion abnormalities in brain diseases \cite{Mer+17,Vil+17}.

	%%%%%%%%%%%%%%%%%%%%%%%%%%%%%%%%%%%%%%%%%%%%%%%%%%%%%%%%%%%%%%%%%
	\section{Theory.}
	
	%%%%%%%%%%%%%%%%%%%%%%%%%%%%%%%%%%%%%%%%%%%%%%%%%%%%%%%%%%%%%%%%%
	\subsection{Forward model.}
	
	The discrete MRI measurement model for a single image with cartesian measurements is of the form
	\begin{equation}
	m = \mathcal{F} u + e,
	\end{equation}
	where $m\in\mathbb{C}^M$ is a complex valued measurement vector, where $M$ is the number of k-space measurement points, $\mathcal{F}$ is the discrete Fourier transform, $u\in\mathbb{C}^N$ is a complex valued (vectorized) image, where $N$ is the number of pixels in the image, and $e$ models measurement noise. In the case of a non-cartesian k-space sampling trajectory, the Fourier transform can be approximated with the non-uniform fast Fourier transform (NUFFT) operation \cite{FS03}.
	
	When NUFFT is used as the forward model, the measurement model can be written as
	\begin{equation}
	m = P\mathcal{F}S u + e,
	\end{equation}
	where $P$ is an interpolation and sampling matrix from the cartesian k-space to the non-cartesian k-space trajectory and $S$ is a scaling matrix. Hereafter we denote $A\coloneqq P\mathcal{F}S$.
	
	When considering dynamic MRI with complementary k-space sampling, where different (undersampled) trajectories of the k-space are measured at different time points, the forward model changes depending on the time point. The forward model can then be written as
	\begin{equation}
	m^t = P^t\mathcal{F}S^t u^t+e^t =A^t u^t+e^t,
	\end{equation}
	where the integer valued superscript $t$ denotes the time index of the measurement and image series, and $m^t$ is the vector of k-space data for a single image in the time series, which is dependent on the selected data segmentation length.

	%%%%%%%%%%%%%%%%%%%%%%%%%%%%%%%%%%%%%%%%%%%%%%%%%%%%%%%%%%%%%%%%%
	\subsection{Joint reconstruction formulation of the dynamic inverse problem.}
	
	The CS based joint image reconstruction formulation we consider is 
	\begin{equation}
	\uest =\argmin_{u=u^1,u^2,...,u^T} \biggl\{ \sum_{t=1}^{T}\Bigl[ \norm{A^tu^t-m^t}_2^2 + \alpha\norm{\nabla_\mathrm{S} u^t}_1\Bigr] + \beta \norm{\nabla_\mathrm{T} u}_1 \biggr\},
	\label{eq:basic_model}
	\end{equation}
	where $T$ is the number of image frames in the problem, $\nabla_\mathrm{S}$ is the discrete spatial gradient operator, $\nabla_\mathrm{T}$ is the discrete temporal gradient operator, and $\alpha$ and $\beta$ are regularization parameters for spatial and temporal regularizations respectively.
	
	Here, we use the isotropic form of 2D spatial TV, where the total variation functional for a complex valued vectorized image $u^t$ is defined as
	\begin{equation}
	\norm{\nabla_\mathrm{S}u^t}_1 =\sum_{k=1}^N \Bigl( (\operatorname{Re}(\mathrm{D}_\mathrm{x}^ku^t))^2+(\operatorname{Re}(\mathrm{D}_\mathrm{y}^ku^t))^2+
	(\operatorname{Im}(\mathrm{D}_\mathrm{x}^ku^t))^2+(\operatorname{Im}(\mathrm{D}_\mathrm{y}^ku^t))^2\Bigr)^{1/2},
	\end{equation}
	where $\operatorname{Re}$ and $\operatorname{Im}$ denote taking the real and the imaginary parts of the complex valued image, $k$ denotes the spatial index in the 2D images, and $D_x^k$ and $D_y^k$ are discrete forward differences in the horizontal and vertical image directions of the $k$'th pixel defined as
	\begin{align}
	\mathrm{D}_\mathrm{x}^ku^t &= 
	\begin{cases*}
	-u_k^t+u_{k+n}^t, & if $k \leq N-n$ \\
	0, & otherwise
	\end{cases*}
	\\
	\mathrm{D}_\mathrm{y}^ku^t &=
	\begin{cases*}
	-u_k^t +u_{k+1}^t, & if $k\Mod{n} \neq 0$\\
	0, & otherwise,
	\end{cases*}
	\end{align}
	where $n$ is the number of rows and columns in the image that is assumed to be square.
	
	Similarly, the temporal TV is defined by
	\begin{equation}
	\norm{\nabla_\mathrm{T}u}_1 = \sum_{t=1}^T\sum_{k=1}^N \sqrt{(\operatorname{Re} (\mathrm{D}_\mathrm{T}^tu_k))^2 +(\operatorname{Im} (\mathrm{D}_\mathrm{T}^tu_k))^2},
	\end{equation}
	where $u_k=u_k^1,...,u_k^T$ and $\mathrm{D}_\mathrm{T}^t$ is the discrete forward difference in the temporal direction of the $t$'th image defined as
	\begin{equation}
	\mathrm{D}_\mathrm{T}^tu_k=
	\begin{cases*}
	-u_k^t+u_k^{t+1}, & if $t \neq T$\\
	0, & otherwise.
	\end{cases*}
	\end{equation}

	%%%%%%%%%%%%%%%%%%%%%%%%%%%%%%%%%%%%%%%%%%%%%%%%%%%%%%%%%%%%%%%%%
	\subsection{Selection of the regularization parameters.}
	
	The S-curve was originally proposed for the selection of a regularization parameter in wavelet based sparsity promoting regularization in \cite{Kol+12}. The idea in the S-curve method is to select the regularization parameter such that the reconstructed image has an \textit{a priori} expected sparsity level in the regularization domain.
	
	The CS formulation of the dynamic MRI problem in \eqref{eq:basic_model} includes two regularization parameters, $\alpha$ and $\beta$, making the parameter selection a two dimensional problem. Here, the idea of the S-curve method is extended to the selection of two regularization parameters, which we propose to select by finding the parameter pair $(\ahat,\bhat)$ that produces expected sparsity level in both the spatial and temporal regularization domains, with simultaneous or sequential selection of the parameters.

	%%%%%%%%%%%%%%%%%%%%%%%%%%%%%%%%%%%%%%%%%%%%%%%%%%%%%%%%%%%%%%%%%
	\subsubsection{Selection of the {\it a priori} sparsity estimates.}
	
	Assume that we have an {\it a  priori} estimate 
	\begin{equation}
	\hattvs = \tvs (\uref) := \norm{\nabla_\mathrm{S} \uref}_1
	\end{equation}
	for the spatial total variation norm of a single unknown image $u^t$, obtained from a reference image $\uref$. The reference image can be for example an anatomical image from fully sampled measurements, a conventional reconstruction of a long sequence of the baseline of the dynamic data or an anatomical atlas image.
	
	Remark that in cases where the expected sparsity level $\hattvs$ is estimated from a reference image $\uref$ that has been acquired with different contrast parameters or comes from a different imaging modality, the reference image may need to be normalized to the scale of the dynamic images. This normalization of the reference image can be done, for example, by
	\begin{equation} \label{normalisation}
	\uref \leftarrow \frac{\norm{m^1}_2}{\norm{A^1\uref }_2} \uref,
	\end{equation}
	where $m^1$ is the first baseline frame of dynamic MRI data and $A^1$ is the respective forward model.
	
	When the temporal changes in the image are based on (mostly) unidirectional contrast changes, an estimate for the sparsity level $\hattvt$ of the temporal gradient can be obtained from the dynamic data using the zero frequency (DC) components of the k-space data. The sparsity estimate is based on the property that the DC component of the Fourier transform of a function $f$ equals the total intensity of the image, i.e.
	\begin{equation}
	\hat{f}(0) = \int_\Omega f(r) {\rm d}r.
	\end{equation}
	Therefore, for the total intensity difference between two images $(w,z)$ we have
	\begin{equation}
	\abs{\int_\Omega  w(r){\rm d}r - \int_\Omega z(r) {\rm d}r}=\abs{\hat{w}(0)-\hat{z}(0)}.
	\end{equation}
	Thus, an estimate for the temporal total variation
	\begin{equation}
	\tvt (u) = \norm{\nabla_\mathrm{T} u}_1
	\end{equation}
	of the unknown image sequence $u$ can be obtained from the measurement data by
	\begin{equation} \label{tvtest}
	\hattvt \approx \sum_{t = 1}^{T-1} \abs{ m_{(k=0)}^{t+1} - m_{(k=0)}^{t} },
	\end{equation}
	where the subscript $(k=0)$ refers to one of the zero frequency components of the k-space data vector $m^t$. 
	
	Note that under ideal noiseless measurement data, the estimate $\hattvt$ would be a lower bound for the temporal TV of the image sequence. The difference between the zero-frequency k-space coefficients of two images would be equal to the temporal TV of the images when there is no noise or tissue movement and the contrast changes between the two images are completely unidirectional. When the contrast changes between the two images are of different signs in different parts of the target or the image changes are caused by motion, the difference of the zero-frequency k-space coefficients would underestimate the temporal TV between the images. In presence of measurement noise or forward model errors, the estimate (\ref{tvtest}) may also be larger than the true temporal TV of the unknown image sequence.

	%%%%%%%%%%%%%%%%%%%%%%%%%%%%%%%%%%%%%%%%%%%%%%%%%%%%%%%%%%%%%%%%%
	\subsubsection{Simultaneous selection of the parameters (S-surface).}
	\label{sec:S-surface}
	
	Given the estimates $\hattvt$ and $\hattvs$ of the \textit{a priori} expected temporal and spatial sparsity, we select the regularization parameters $(\alpha,\beta)$ as follows:
	\begin{itemize}
		\item [1)] Take a grid of regularization parameters $\alpha$ and $\beta$ ranging on the interval $(0, \infty)$ such that
		$$
		0 < \beta^{(1)} < \beta^{(2)} < \cdots < \beta^{(P)} < \infty 
		$$
		and
		$$
		0 < \alpha^{(1)} < \alpha^{(2)} < \cdots < \alpha^{(L)} < \infty .
		$$
		\item [2)] Compute the corresponding estimates $\uest (\alpha^{(\ell)},\beta^{(p)}),
		\ell = 1, \ldots, L,\ p=1,\ldots,P$. 
		The reconstructions $\uest (\alpha^{(\ell)},\beta^{(p)}) $ are computed by
		\begin{equation}
		\uest =\argmin_{u=u^1,u^2,...,u^T} \biggl\{ \sum_{t=1}^{T}\Bigl[ \norm{A^tu^t-m^t}_2^2 + 
		\alpha^{(\ell)} \norm{\nabla_\mathrm{S} u^t}_1\Bigr] + \beta^{(p)} \norm{\nabla_\mathrm{T} u}_1 \biggr\}.
		\label{eq:basic_model_alpha_beta}
		\end{equation}
		Here, it is crucial that $\beta^{(1)}$ is selected so small that the reconstructions have a larger temporal TV than the expected temporal sparsity level, and $\alpha^{(1)}$ is so small that the reconstructions have a larger spatial TV than the expected spatial sparsity level. Similarly, the values $\beta^{(P)}$ and $\alpha^{(L)}$ are selected so large that the problem becomes over-regularized and the respective TV values of the reconstructions are small. 
		\item[3)] Compute the temporal TV of the image series $\tvt (\uest (\alpha,\beta))$ and the spatial TV of one time frame of the image series $\tvs (\uest^t (\alpha,\beta))$ for each grid point $(\alpha^{(\ell)},\beta^{(p)})$.
		\item [4)] Evaluate the value of the merit functional
		\begin{equation}
		\Psi(\alpha,\beta) = \frac{\abs{\tvt (\uest (\alpha,\beta))-\hattvt}}{2\hattvt} + \frac{\abs{\tvs (\uest^t (\alpha,\beta))-\hattvs}}{2\hattvs} 
		\end{equation}
		for each grid point.
		\item[5)] Select the parameter pair $(\ahat,\bhat)$ which minimizes $\Psi(\alpha,\beta)$. 
	\end{itemize}

	%%%%%%%%%%%%%%%%%%%%%%%%%%%%%%%%%%%%%%%%%%%%%%%%%%%%%%%%%%%%%%%%%
	\subsubsection{Sequential selection of the parameters (S-curve).}
	\label{ssec:s-curve}
	
	The two-dimensional selection of the parameters $(\alpha,\beta)$ in Section \ref{sec:S-surface} requires computing a large number of reconstructions of the form of  \eqref{eq:basic_model} over a 2D grid of regularization parameter values, making it computationally expensive, especially in large 4D problems. Therefore, to alleviate the computational burden, we also consider an approach where we split the parameter search into two 1D problems.
	
	We employ the S-curve first for the selection of the temporal regularization parameter $\beta$ without any spatial regularization (i.e. $\alpha = 0$ in \eqref{eq:basic_model}), and then we employ the S-curve for the selection of the spatial regularization parameter $\alpha$ using the value of $\beta$ found in the first step. We select the parameter $\beta$ first since the reconstruction from undersampled measurements is more dependent on the temporal regularization than the spatial regularization.
	
	{\bf Selection of $\beta$:} Given the estimate $\hattvt$ of the \textit{a priori} temporal sparsity level, we first select the temporal regularization parameter $\beta$ using the S-curve as follows:
	\begin{itemize}
		\item [1)] Take a sequence of the temporal regularization parameters $\beta$ ranging on the interval $(0, \infty)$ such that
		$$
		0 < \beta^{(1)} < \beta^{(2)} < \cdots < \beta^{(P)} < \infty .
		$$
		\item [2)] Compute the corresponding estimates $\uest (\beta^{(1)}),
		\ldots, \uest (\beta^{(P)})$ according to
		\begin{equation}
		\uest (\beta^{(p)}) =\argmin_{u=u^1,u^2,...,u^T} \biggl\{ \sum_{t=1}^{T} \norm{A^tu^t-m^t}^2  + \beta^{(p)} \norm{\nabla_\mathrm{T} u}_1 \biggr\}.
		\label{eq:basic_model_beta}
		\end{equation}
		Here too, $\beta^{(1)}$ has to be so small that the problem is under-regularized and the corresponding reconstruction $\uest (\beta^{(1)})$ is a noisy image sequence with a temporal TV larger than $\hattvt$ and $\beta^{(P)}$ is so large that the problem is over-regularized and the temporal TV of the reconstruction is close to zero.
		\item[3)] Compute the temporal TV of the recovered estimates $\uest (\beta^{(p)}), \quad p = 1,\ldots,P$.
		\item [4)] Fit a smooth interpolation curve to the data $\{ \beta^{(p)},\tvt(\uest (\beta^{(p)})), \quad p = 1,\ldots,P\}$ and use the interpolated sparsity curve to find the value  $\bhat$ for which $\tvt (\bhat) = \hattvt$. 
	\end{itemize}
	
	{\bf Selection of $\alpha$:} Given the value $\bhat$ and the \textit{a priori} expected spatial sparsity level $\hattvs$, the spatial regularization parameter $\alpha$ is selected according to the following procedure: 
	\begin{itemize}
		\item [1)] Take a sequence of the spatial regularization parameters $\alpha$ ranging on the interval $(0, \infty)$ such that
		$$
		0 < \alpha^{(1)} < \alpha^{(2)} < \cdots < \alpha^{(L)} < \infty .
		$$
		\item [2)] Compute the corresponding estimates $\uest (\alpha^{(1)}),
		\ldots, \uest (\alpha^{(L)})$ by
		\begin{equation}
		\uest (\alpha^{(\ell)},\bhat) =\argmin_{u=u^1,u^2,...,u^T} \biggl\{ \sum_{t=1}^{T}\Big[ \norm{A^tu^t-m^t}_2^2 + \alpha^{(\ell)} \norm{\nabla_\mathrm{S} u^t}_1\Big] + \bhat \norm{\nabla_\mathrm{T} u}_1 \biggr\}.
		\label{eq:basic_model_alpha}
		\end{equation}
		Again, $\alpha^{(1)}$ needs to be so small that a frame in the reconstruction $\uest (\alpha^{(1)})$ results in a spatial TV larger than $\hattvs$ and $\alpha^{(L)}$ needs to be so large that the spatial TV of a frame  $\uest (\alpha^{(L)})$ is very small.
		\item[3)] Compute the spatial TV of a frame of the recovered estimates $\uest (\alpha^{(\ell)},\bhat), \quad \ell = 1,\ldots,L$, i.e. $\tvs(\uest^t (\alpha^\ell,\bhat))$.
		\item [4)] Fit a smooth
		interpolation curve to the data $\{ \alpha^{(\ell)},\tvs(\uest^t(\alpha^{(\ell)},\bhat)), \quad \ell = 1,\ldots,L\}$, and use the interpolated sparsity curve to find the value  $\ahat$ for which $\tvs(\ahat,\bhat) = \hattvs$.
	\end{itemize}
	The final reconstruction for analysis and diagnostic tasks can then be computed using the parameter pair $(\ahat,\bhat)$.

	%%%%%%%%%%%%%%%%%%%%%%%%%%%%%%%%%%%%%%%%%%%%%%%%%%%%%%%%%%%%%%%%%
	\subsubsection{L-curve parameter selection.}
	The L-curve parameter selection \cite{HO93} was computed as a reference for the proposed sparsity based methods. Similarly to the S-curve method, the L-curve was implemented sequentially such that the temporal regularization parameter $\beta$ was selected first, and then the spatial regularization parameter $\alpha$ was selected using a fixed $\beta$.
	
	{\bf Selection of regularization parameters:} To compute the ensemble of reconstructions with different $\beta$:s or $\alpha$:s, steps 1) and 2) of the respective S-curve parameter selection in section \ref{ssec:s-curve} were first computed. After this the L-curve method was used as follows: 
	\begin{itemize}
		\item [3)] Set $\rho=\log_{10}\sum_{t=1}^{T} \norm{A^t u^t-m^t}_2^2$, $\eta=\log_{10}\norm{R(u)}$, where, $R(u)$ is either the sum of the spatial total variations of all the frames $\sum_{t=1}^T\norm{\nabla_\mathrm{S}u^t}_1$ or the temporal total variation $\norm{\nabla_\mathrm{T}u}_1$.
		\item [4)] Interpolate the L-curve ($\rho$,$\eta$) to a dense grid using spline interpolation.
		\item [5)] Calculate the maximum curvature point of the L-curve using
		\begin{equation}
		\lambda=\arg\max\kappa(\lambda)=\arg\max\frac{\rho'\eta''-\rho''\eta'}{((\rho')^2+(\eta')^2)^{3/2}},
		\end{equation}
		where $\lambda$ is either $\alpha$ or $\beta$.
	\end{itemize}

	%%%%%%%%%%%%%%%%%%%%%%%%%%%%%%%%%%%%%%%%%%%%%%%%%%%%%%%%%%%%%%%%%
	\subsubsection{Monte-Carlo SURE parameter selection.}
	Monte-Carlo SURE (MC-SURE) parameter selection \cite{RBU08,Ram+13,Wel+14} was also computed as a reference. MC-SURE estimates a weighted mean squared-error using the principles of Stein's unbiased risk estimate \cite{Ste81} by perturbing the data vector with a complex-valued random vector and analyzing the response to the perturbation. Similar to both the S-curve method and the L-curve method, the MC-SURE parameter selection was first done for the temporal regularization parameter $\beta$ and then the spatial regularization parameter $\alpha$ with a fixed $\beta$.
	
	{\bf Selection of regularization parameters:}
	The full MC-SURE function is of the form
	\begin{equation}
	\label{eq:mcsure_full}
	\textrm{MC-SURE}(\lambda) = M^{-1}\norm{Au_\lambda(m)-m}_W^2-M^{-1}\mathrm{tr}\{\Omega W\}+2M^{-1}\operatorname{Re}\{\mathrm{tr}\{\Gamma J_{u_\lambda(m)}\}\},
	\end{equation}
	where $\lambda$ is the parameter pair $(\alpha,\beta)$, $u_\lambda(m)$ is the reconstruction computed with the regularization parameter pair $\lambda$ and measurements $m$, $W$ is a weighting matrix, $\Omega$ is the noise covariance matrix, $\Gamma = \Omega W A$ and $M$ is the number of measurement points. Further, the second trace is approximated by
	\begin{equation}
	\mathrm{tr}\{\Gamma J_{u_\lambda(m)}\}\approx \epsilon^{-1}b'\Lambda^{-1}\Gamma\rho(u_\lambda,m,\Lambda b,\epsilon),
	\end{equation}
	where $\epsilon$ is a data perturbation multiplier, $\Lambda$ is a matrix allowing different amounts of perturbation for different elements of $m$, $b$ is a complex valued random vector and $\rho$ is the perturbation between the reconstructions from the perturbed data and the original data
	\begin{equation}
	\rho(u_\lambda,m,\Lambda b,\epsilon) = u_\lambda(m+\epsilon\Lambda b) - u_\lambda(m).
	\end{equation}
	
	Here, as in \cite{Ram+13}, we set $b=(b_{\operatorname{Re}}+ib_{\operatorname{Im}})/\sqrt{2}$, where $b_{\operatorname{Re}}$ and $b_{\operatorname{Im}}$ are independent binary random vectors of -1 and 1 with equal probability. In addition, we set $\Lambda=I$ and $W=I$. Since we are minimizing \eqref{eq:mcsure_full} w.r.t. $\lambda=(0,\beta)$ or $\lambda=(\alpha,\hat{\beta})$, where $\hat{\beta}$ is fixed, we can in addition leave out the second term, which is the same for all $\lambda$, and the multiplier $M^{-1}$. The approximated and shortened form of the MC-SURE function is thus
	\begin{equation}
	\label{eq:MC-SURE_final}
	\textrm{MC-SURE}(\lambda) \approx \norm{Au_\lambda(m)-m}^2 + 2\operatorname{Re}\{ \epsilon^{-1}b'\Omega A (u_\lambda(m+\epsilon b) - u_\lambda(m)) \}.
	\end{equation}
	
	In practice, MC-SURE is implemented as follows:
	\begin{itemize}
		\item [1)] Compute the reconstructions \eqref{eq:basic_model_beta} with the original data and the perturbed data for parameter pairs $\lambda=(\alpha=0,\beta^{(p)})$ with $p=1,...,P$.
		\item [2)] Calculate MC-SURE according to \eqref{eq:MC-SURE_final} and choose the parameter $\hat{\beta}$ corresponding to the minimum MC-SURE.
		\item [3)] Compute the reconstructions \eqref{eq:basic_model_alpha} with the original data and the perturbed data for parameter pairs $\lambda=(\alpha^{(\ell)},\hat{\beta})$ with $\ell=1,...,L$.
		\item [4)] Calculate MC-SURE according to \eqref{eq:MC-SURE_final} and choose the parameter $\hat{\alpha}$ corresponding to the minimum MC-SURE.
	\end{itemize}
	The parameter selection by MC-SURE thus requires computing a total of $2(P+L)$ reconstructions.
	
	%%%%%%%%%%%%%%%%%%%%%%%%%%%%%%%%%%%%%%%%%%%%%%%%%%%%%%%%%%%%%%%%%
	\section{Methods.}
	
	%%%%%%%%%%%%%%%%%%%%%%%%%%%%%%%%%%%%%%%%%%%%%%%%%%%%%%%%%%%%%%%%%
	\subsection{Simulated test case.}
	
\begin{figure}
		\centering
		\includegraphics[width=.85\columnwidth]{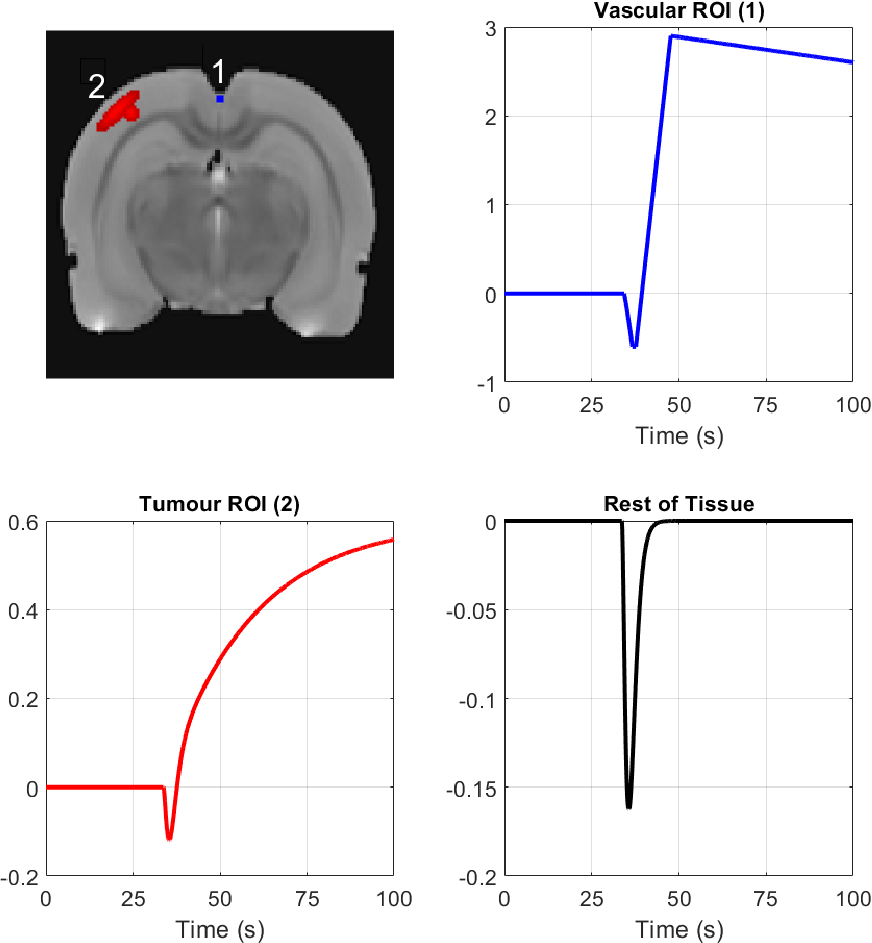}
		\caption{Vascular and tumour regions and the three template signals used in the simulation. Top left: The simulated image with the vascular ROI marked in blue and labelled '1', and the tumour ROI marked in red and labelled '2'. Top right: Simulated vascular ROI signal template. Bottom left: Simulated tumour ROI signal template. Bottom right: Simulated signal template in the rest of the tissue. Each image pixel was multiplied with the corresponding template and the result was added to the image to create the simulated time series.} 
		\label{fig:simu}
	\end{figure}

	A simulation modelling DCE-MRI of a glioma in a rat brain was generated. The rat brain image used as the basis for the simulation was the rat brain atlas image \cite{Val+11}, and the image was scaled to be of size 128x128. The rat brain tissue area was split into three subdomains with varying signal behaviour: 1) vascular region corresponding to the location of the superior sagittal sinus (highlighted with blue and labelled '1' in Fig.~\ref{fig:simu}), 2) generated tumour region (highlighted with red and labelled '2' in Fig.~\ref{fig:simu}) and 3) the rest of the brain tissue. The superior sagittal sinus can be used as a proxy for estimating the arterial input function (AIF) in DCE-MRI of the brain \cite{LV10}. The AIF is used in the estimation of pharmacokinetic parameters in DCE-MRI \cite{Tof+99}, and the vascular region is thus of special interest.
	
	Fig.~\ref{fig:simu} shows the signal templates generated for each of the three regions of interest. 2800 ground truth images were created based on the signal templates by multiplying the signal of each pixel in the original image with the signal template of the corresponding region and adding the result of the multiplication to the original value of the pixel. The three template signals were based on an experimental DCE-MRI measurement, which is described briefly in Sect.~\ref{ssec:invivomeas} and also in \cite{Han+19,Ras+18}, from which the three different ROIs were identified. The same simulated test case has been used previously in \cite{Han+19,Nii+19}.
	
	The simulated test case was carried out using a k-space trajectory which combines the golden angle (GA) \cite{Win+07} and the concentric squares sampling strategies \cite{ACD+06,YLL+16}.
	
	For each of the 2800 ground truth images, one spoke of k-space data were generated. The repetition time of the experimental measurements was 38.5 ms, and the signal dynamics of the simulation were set to reflect this. Finally, complex Gaussian noise at 2\%, 5\% and 10\% of the mean of the absolute values of the signal without noise was added to the simulated k-space signal to create data realizations with different noise levels.

	%%%%%%%%%%%%%%%%%%%%%%%%%%%%%%%%%%%%%%%%%%%%%%%%%%%%%%%%%%%%%%%%%
	\subsubsection{Error metric for the simulation.}
	
	Root mean square error (RMSE) values were calculated for the three regions of interest: 1) vascular region, 2) tumour region and 3) the rest of the image. For the calculation of the RMSE, the
	reconstructed signals of each pixel were linearly interpolated in the temporal direction to match with the temporal resolution of the ground truth phantom (i.e., segment length of one spoke). After the interpolation, RMSE to the ground truth phantom was calculated for all regions separately. After the RMSEs for the three ROIs were computed, a joint RMSE was computed by taking the norm of the separate RMSEs. This was done to weigh the ROIs with a small support (vascular, tumour) appropriately in the error metric. The error metric was selected independent of the parameter selection, and the same metric was used in \cite{Han+19,Nii+19}. The error metric is described in more detail in \cite{Han+19}.

	%%%%%%%%%%%%%%%%%%%%%%%%%%%%%%%%%%%%%%%%%%%%%%%%%%%%%%%%%%%%%%%%%
	\subsection{In vivo measurements from a rat glioma model.}\label{ssec:invivomeas}
	
	The animal experiment was approved by the Animal Health Welfare and Ethics Committee of University of Eastern Finland. DCE-MRI data were acquired from a female Wistar rat with a glioma model \cite{Niv+20}. The experiment was performed 10 days post-implantation of the glioma cells into the rat brain. During the imaging, the animal was anesthetized and kept in fixed position in a holder which was inserted into the magnet. A needle was placed into the animal's tail vein for the injection of the contrast agent.
	
	The DCE data were collected using a 9.4 T horizontal magnet interfaced to an Agilent imaging console and a volume coil transmit/quadrature surface coil receive pair (Rapid Biomed, Rimpar, Germany). The DCE-MRI data were measured with radial golden angle sampling using a gradient-echo based radial pulse sequence with field-of view 32 mm x 32 mm, slice thickness 1.5 mm, repetition time 38.5 ms, echo time 9 ms, flip angle 30 degrees and 128 points in each spoke. 610 spokes were collected in sequential order, after which the next spoke would differ by less than 0.1 degrees from the first spoke, so the cycle of 610 spokes was repeated to simplify the sequence. This cycle was repeated for a total of 25 times, which leads to a sequence of 15250 spokes of data for a measurement duration of nearly 10 minutes. The measurement time for a single cycle of 610 spokes was $610 \cdot 38.5$ ms $= 23.46$ s. One minute after the beginning of the dynamic sequence, Gadovist (1 mmol/kg) was injected i.v. over a period of 3 s. 7320 spokes of data from the beginning of the measurements were used for testing the proposed parameter selection approach in the experimental data results section.
	
\begin{figure}
		\centering
		\includegraphics[width=.85\columnwidth]{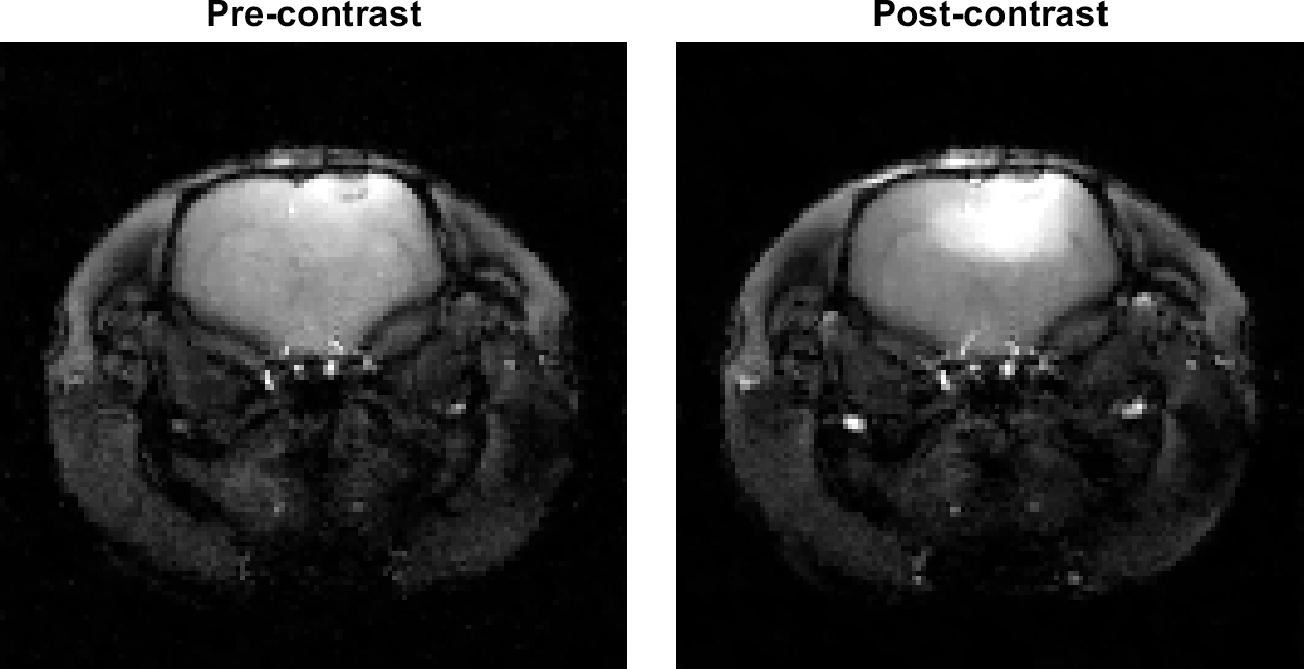}
		\caption{Cartesian gradient-echo pulse sequence full data reconstructions with spatial TV regularization from before and after contrast injection used as reference. The two images have the same adjusted color scale.}
		\label{fig:sems_ref}
	\end{figure}

	Anatomical reference images were also scanned from the same slice before and after the dynamic imaging using a gradient-echo pulse sequence with similar parameters with the dynamic data sequence but using a full Cartesian sampling of 128x128 points of k-space data. Anatomical images reconstructed from the full Cartesian data with spatial regularization from before and after the experiment are shown as reference to the dynamical reconstructions in Fig.~\ref{fig:sems_ref}.
	
	The dynamical experiment was also used as a basis for creating the three signal templates shown in Fig.~\ref{fig:simu} for the simulated test case. For the simulation, three regions were identified from the experimental data reconstructions: vascular region (superior sagittal sinus), glioma region, and the rest of the brain tissue.

	%%%%%%%%%%%%%%%%%%%%%%%%%%%%%%%%%%%%%%%%%%%%%%%%%%%%%%%%%%%%%%%%%
	\subsection{Computation.}
	The Chambolle-Pock primal-dual algorithm \cite{CP11,SJP12} was used to solve the minimization problem of \eqref{eq:basic_model}. The ratio of the primal and dual step sizes was varied according to the regularization coefficients such that the primal step size was smaller (and accordingly the dual step size was larger) for larger regularization parameters. Asymmetrical primal and dual step sizes in the algorithm have been shown to lead to faster convergence in some cases in both linear \cite{PC11} and non-linear \cite{Val14} problems, and this was observed here as well. The operator norms of the forward problem, and the spatial and temporal total variation terms were all scaled to 1.

	%%%%%%%%%%%%%%%%%%%%%%%%%%%%%%%%%%%%%%%%%%%%%%%%%%%%%%%%%%%%%%%%%
	\section{Results.}
	
	%%%%%%%%%%%%%%%%%%%%%%%%%%%%%%%%%%%%%%%%%%%%%%%%%%%%%%%%%%%%%%%%%
	\subsection{Simulated data.}
	For both the simulated and experimental test case, we use a segment length of 34, i.e. each image frame is computed from 34 spokes of GA data. This was found to be an optimal segment length for a similar dynamic simulation in \cite{Han+19}.
	
	For the simulation, we use the parameter pair and the reconstruction that yields the lowest joint RMSE as references of the best possible parameter pair and reconstruction. We refer to this parameter selection method as MinRMSE and compare the S-surface, S-curve, L-curve and MC-SURE methods to the MinRMSE. The regularization parameters selected with the different methods for the simulation with 5\% noise are shown in Table \ref{tab:parameters}.
	\begin{table}
		\caption{The spatial ($\ahat$) and temporal ($\bhat$) regularization parameters for the simulated test case with 5\% noise level obtained with the S-surface, S-curve, L-curve and MC-SURE methods and the MinRMSE parameters used as reference.}
		\label{tab:parameters}
		\begin{center}
			\begin{tabular}{ c|c|c|c|c|c }
				& MinRMSE & S-surface & S-curve & L-curve & MC-SURE \\ 
				\hline
				$\ahat$ & 0.0001 & 0.00032 & 0.00019 & 8.8e-5 & 0.0001 \\
				$\bhat$ & 0.0032 & 0.0032 & 0.0046 & 2.7e-5 & 0.001
			\end{tabular}
		\end{center}
	\end{table}
	
	The reference temporal sparsity for the S-surface and S-curve methods was obtained by taking the spoke closest to vertical from each segment of 34 spokes, and calculating the total intensity changes using the DC signal of those spokes. This selection was done due to the DC signal of the real measurements having some dependence on the angle of the measurement spoke. The reference spatial sparsity level used was the spatial total variation of the first frame of the ground truth phantom. Note here, that since the temporal regularization parameter was first selected, the selection of the spatial regularization parameter value is not as sensitive as the temporal regularization parameter selection.
	
	For MC-SURE parameter selection, $\epsilon$ was set to $10^{-3}$ and the noise covariance $\Omega$ was set to the scalar noise level $\sigma^2$, which was obtained from the sample variance of both ends of all the measurement spokes.
	
\begin{figure}
		\centering
		\includegraphics[width=\textwidth]{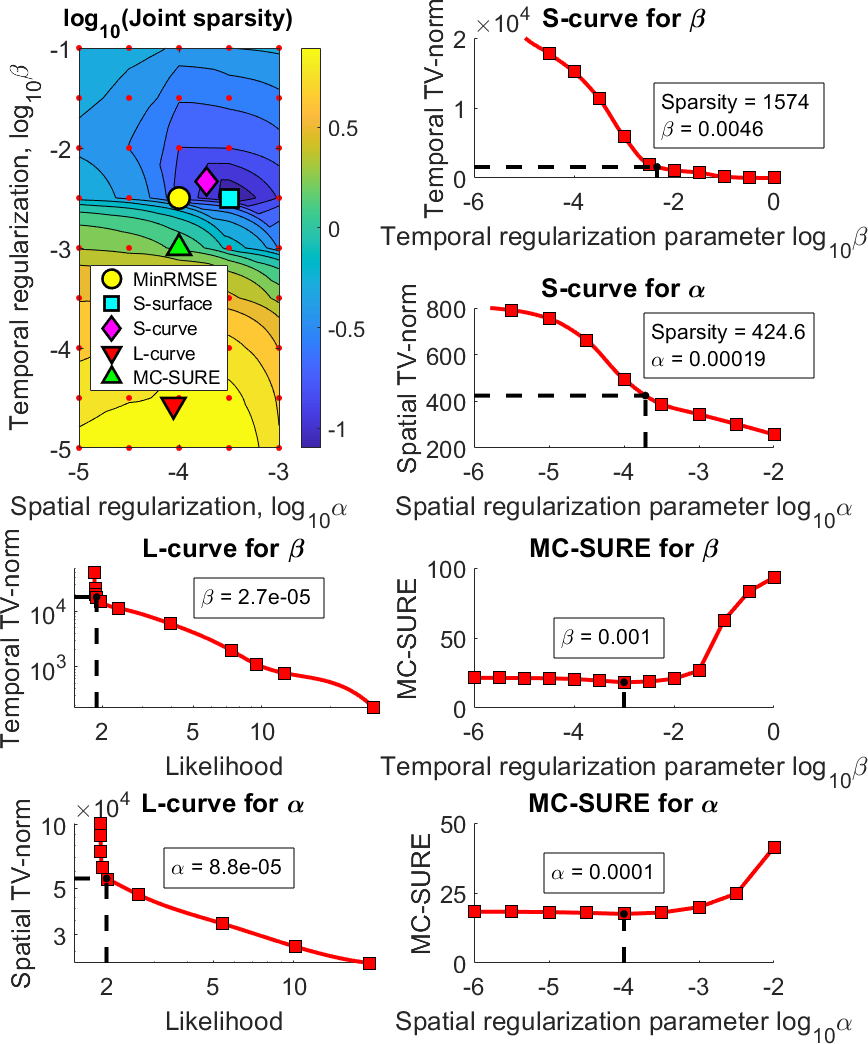}
		\caption{The joint sparsity contour used in the S-surface parameter selection as well as all the parameter selection curves for S-curve, L-curve and MC-SURE in the simulation with noise level 5\%. The red dots mark the computational grid of $\alpha$:s and $\beta$:s in the contour, and the red squares mark the 1D grid of regularization parameters in the parameter selection curves. Note that the spatial TV norm for the S-curve is from a single image frame whereas the spatial TV norm for the L-curve is from the whole image series.}
		\label{fig:simures1}
	\end{figure}

\begin{figure}
		\centering
		\includegraphics[width=\textwidth]{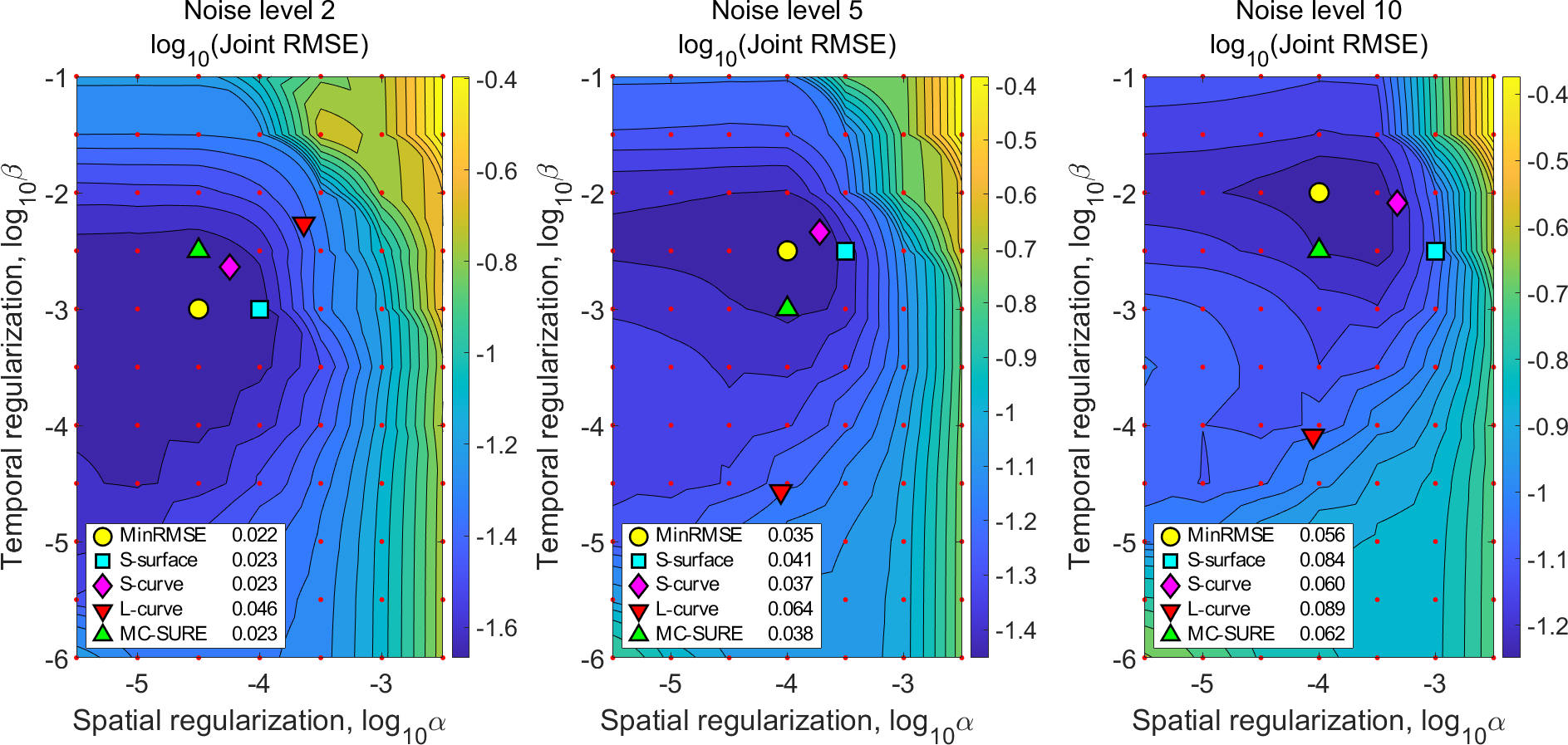}
		\caption{Joint RMSE contours with noise levels 2\%, 5\% and 10\% with the parameters of the four different reconstructions marked and the jRMSE values of the chosen reconstructions. The images show that the regularization parameter pairs obtained with the sparsity based S-surface and S-curve methods are close to the MinRMSE parameter pair.}
		\label{fig:simures2}
	\end{figure}

\begin{figure*}
		\centering
		\subfloat{\includegraphics[width = .9\textwidth]{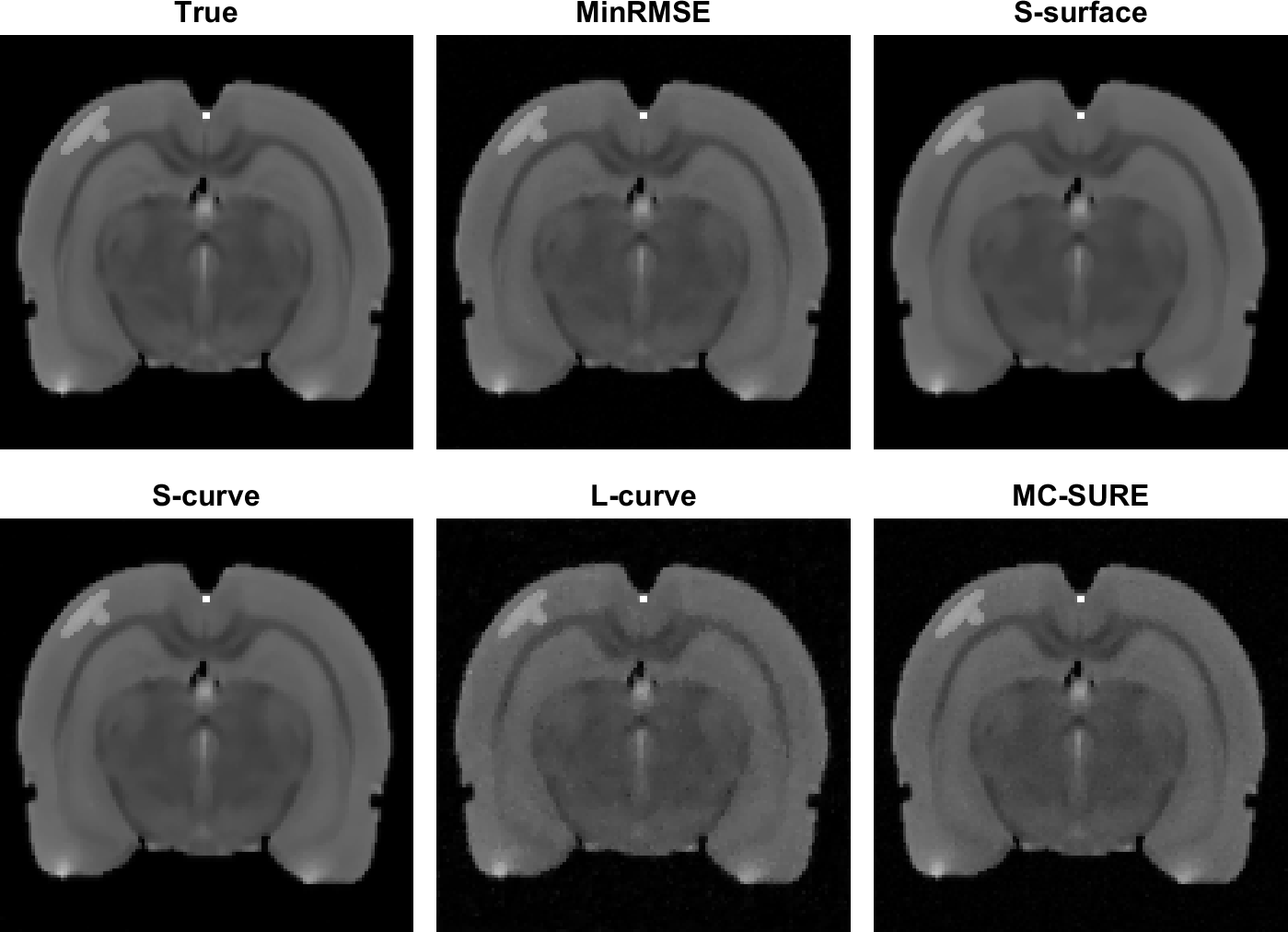}}\\
		\vspace{0.5cm}
		\subfloat{\includegraphics[width = \textwidth]{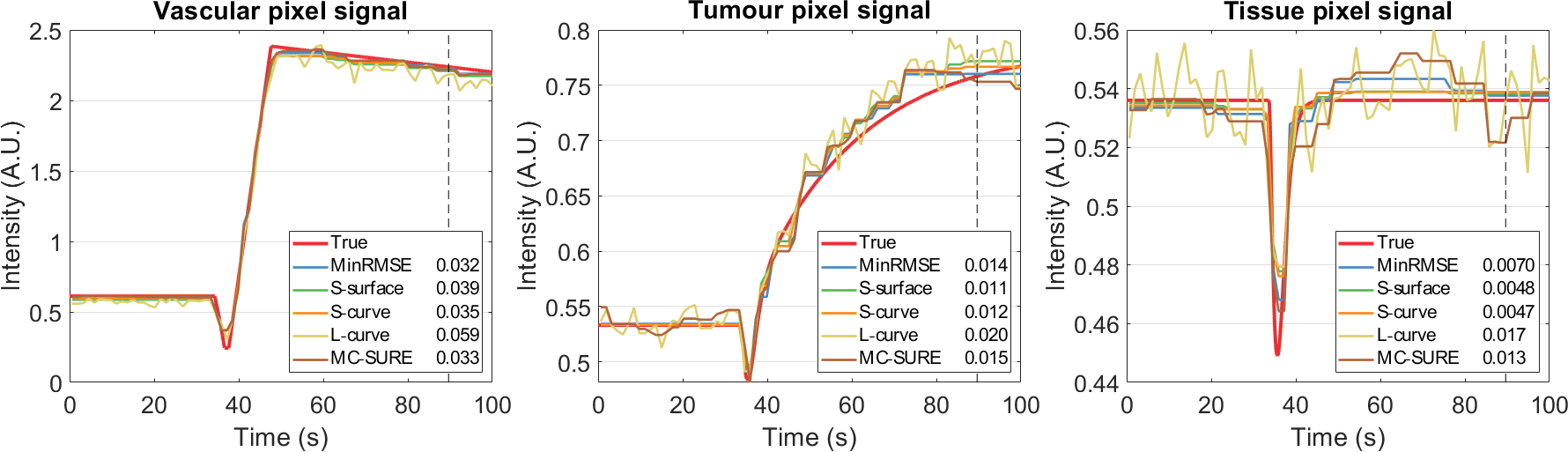}}
		\caption{Top rows: Single frame images of the simulated true target and reconstructions of the simulation with noise level 5\% at $t\approx 90$ s (marked in the signal curves with a dashed line) with parameters chosen with different methods; minimum joint RMSE (MinRMSE), simultaneous parameter selection according to joint sparsity (S-surface), sequential parameter selection according to sparsity (S-curve) and sequential parameter selection by the L-curve and MC-SURE methods. Bottom row: Single pixel signals from the vascular and tumour areas as well as healthy tissue from the right side of the cortex with the different methods compared to the phantom and RMS errors of the corresponding regions.}
		\label{fig:simures3}
	\end{figure*}

	Figs.~\ref{fig:simures1} - \ref{fig:simures3} show the results using the simulated rat brain DCE data. Fig.~\ref{fig:simures1} shows the joint sparsity contour used to select the S-surface parameter pair as well as the parameter selection curves for the S-curve, L-curve and MC-SURE methods in the simulated case with 5\% noise.
	
	Fig.~\ref{fig:simures2} shows the joint RMSE contours of the reconstructions calculated on a wide grid of the spatial and temporal regularization parameters for the simulations with noise levels 2\%, 5\% and 10\%. Fig.~\ref{fig:simures2} also shows the joint RMSE values of the reconstructions with the parameters chosen with the three different methods. The joint RMS error values for the S-curve, S-surface and MC-SURE reconstructions are similar to each other with the noise levels 2\% and 5\% and close to those of the optimal reconstruction (MinRMSE), and the regularization parameters obtained with these methods are also close to the MinRMSE parameters. In the simulation with 10\% noise level, the S-surface performs worse than S-curve and MC-SURE in the joint RMSE measure. The L-curve method performs worse than the other methods with all three noise levels.
	
	Fig.~\ref{fig:simures3} shows single frames of the reconstructions of the data with 5\% noise after contrast agent injection as well as single pixel signals from the tumour and vascular areas as well as healthy tissue on the right side of the cortex and the RMSEs of the three regions. The MinRMSE, S-surface and S-curve reconstructions (with 5\% noise level) in Fig. \ref{fig:simures2} appear mostly similar, whereas the L-curve and MC-SURE reconstructions are slightly noisier. Fig.~\ref{fig:simures3} also shows the RMS errors of the three different regions. In the region based RMS error measures, the S-curve and S-surface yield the best accuracy in the tumour and tissue signals, while MC-SURE has the best accuracy in the vascular region.
	
	{}
	
	{}
	
	{}

	%%%%%%%%%%%%%%%%%%%%%%%%%%%%%%%%%%%%%%%%%%%%%%%%%%%%%%%%%%%%%%%%%
	\subsection{Experimental data.}
\begin{figure}
		\centering
		\includegraphics[width=\textwidth]{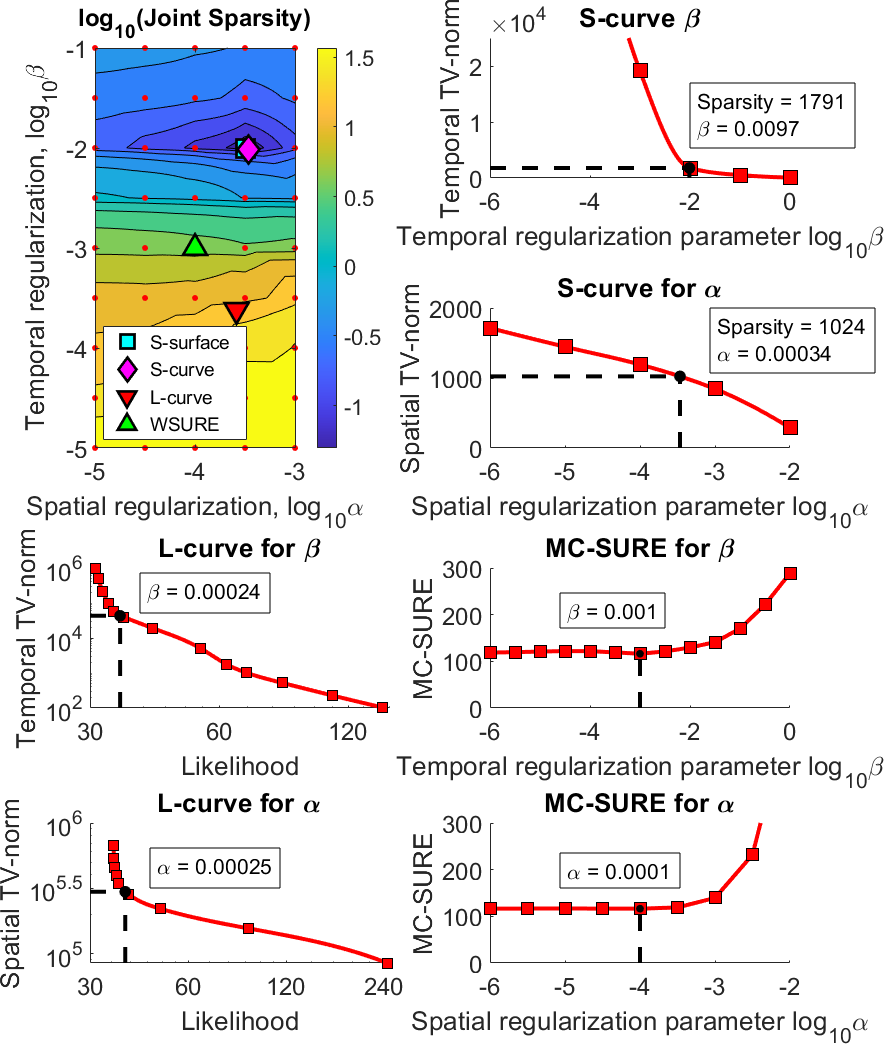}
		\caption{The joint sparsity contour used in the S-surface parameter selection as well as all the parameter selection curves for the S-curve, L-curve and MC-SURE methods in the experimental data case. The red dots mark the computational grid of $\alpha$:s and $\beta$:s in the contour, and the red squares mark the 1D grid of regularization parameters in the parameter selection curves. Note that the spatial TV norm for the S-curve is from a single image frame whereas the spatial TV norm for the L-curve is from the whole image series.}
		\label{fig:experes1}
	\end{figure}

\begin{figure}
		\centering
		\subfloat{\includegraphics[width=.9\textwidth]{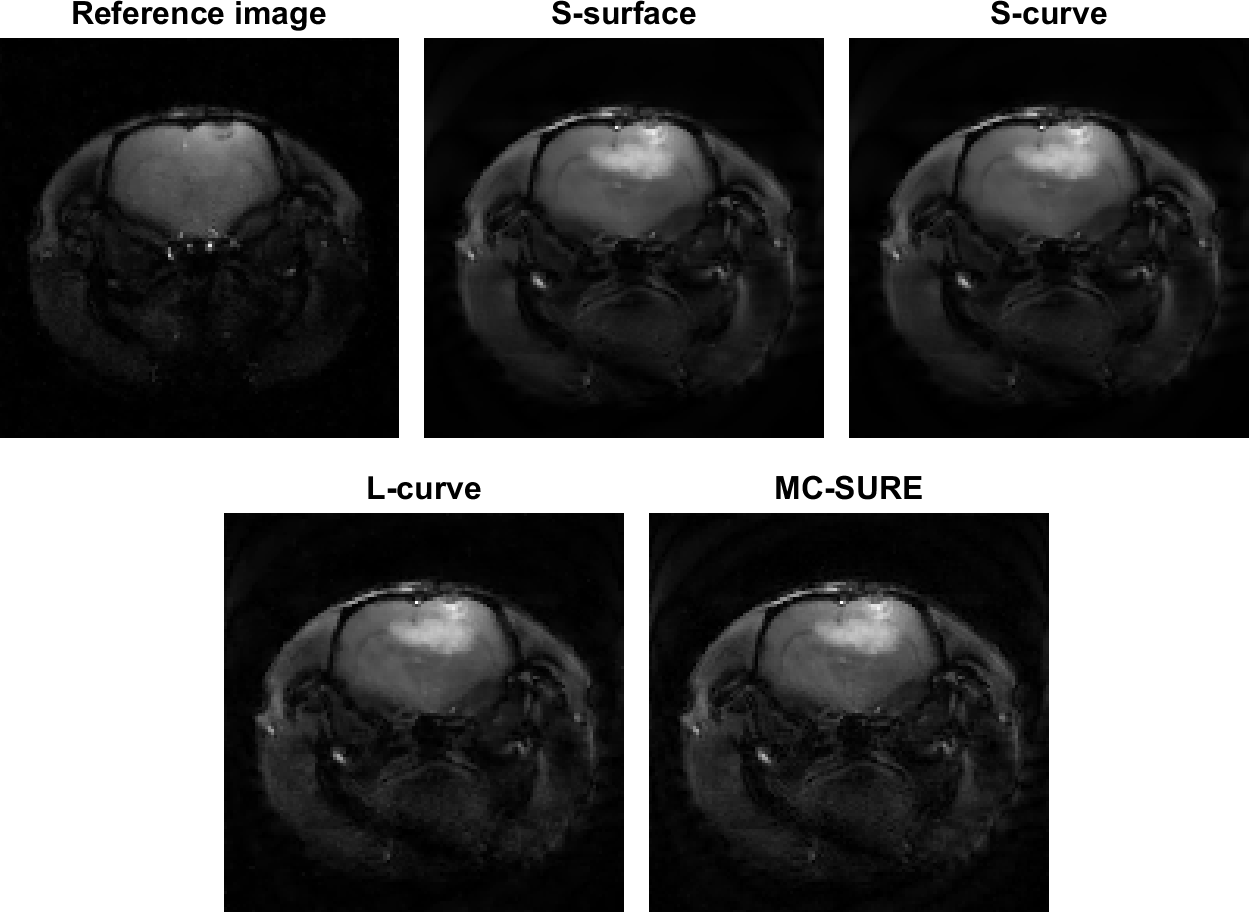}}\\
		\vspace{0.5cm}
		\subfloat{\includegraphics[width = \textwidth]{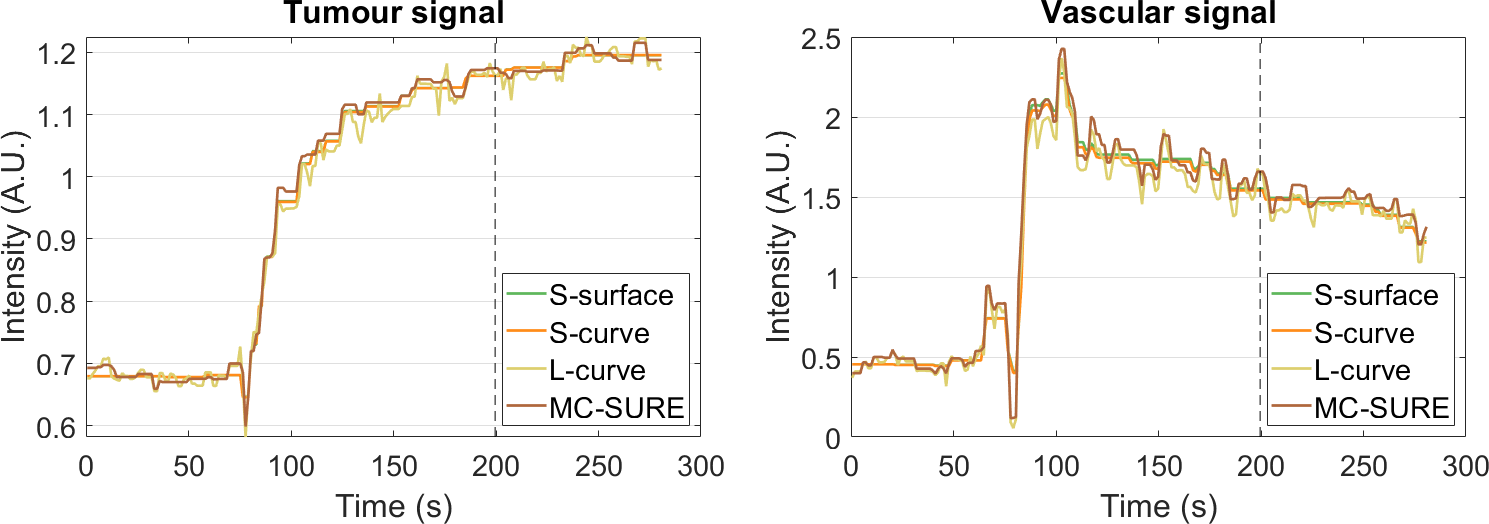}}
		\caption{Top rows: Single frame images of the cartesian reference image with spatial regularization from before contrast agent injection and the reconstructions of the experimental data at $t\approx 200$ s (marked below with a dashed line) with parameters chosen with the S-surface, S-curve, L-curve and MC-SURE methods. Bottom row: Single pixel signals from the tumour and vascular areas with the four different methods.}
		\label{fig:experes2}
	\end{figure}

	Figs.~\ref{fig:experes1} - \ref{fig:experes2} show the results for the experimental data. Fig.~\ref{fig:experes1} shows the joint sparsity grid of reconstructions with a large grid of spatial and temporal regularization parameters where the S-surface, S-curve, L-curve and MC-SURE parameter pairs are marked. The figure also shows all the parameter selection curves used to select both $\alpha$ and $\beta$ with the S-curve, L-curve and MC-SURE methods. The parameters obtained with the S-surface and S-curve methods are almost the same, namely, the S-curve parameters are $\alpha=0.00034$, $\beta=0.0097$, and the S-surface parameters are $\alpha=0.00032$, $\beta=0.01$. The parameters obtained with the L-curve method are $\alpha=0.00025$, $\beta=0.00024$ and the parameters obtained with the MC-SURE method are $\alpha = 0.0001$, $\beta = 0.001$.
	
	For MC-SURE parameter selection, similar to the simulation, $\epsilon$ was set to $10^{-3}$ and the noise covariance $\Omega$ was set to the scalar noise level $\sigma^2$, which was obtained from the sample variance of both ends of all the measurement spokes.
	
	Fig.~\ref{fig:experes2} shows the cartesian reference reconstruction with spatial regularization from before contrast agent injection and single frame images from the dynamic reconstructions using parameters obtained with the different methods from after the contrast agent injection. The reference image is the same as in Fig. \ref{fig:sems_ref} with different contrast. Fig.~\ref{fig:experes2} also contains single pixel signals from the tumour and vascular regions. The reconstructions are mostly visually similar, and the single pixel signals of the S-curve and S-surface reconstructions show less variation than those of the L-curve and MC-SURE reconstructions which yield smaller values of the temporal regularization parameter $\beta$ than the sparsity based selections.
	
	{}
	
	{}

	%%%%%%%%%%%%%%%%%%%%%%%%%%%%%%%%%%%%%%%%%%%%%%%%%%%%%%%%%%%%%%%%%
	\section{Discussion.}
	Compressed sensing based models for dynamic MRI problems usually include regularization for sparsity in both the spatial and temporal domain. Therefore these models typically include two regularization parameters, one for the spatial and one for the temporal regularization, that the user has to select. Often, the selection of both of the parameters is carried out manually based on visual assessment of the reconstructed images.
	
	In this work, we investigated a two-dimensional parameter selection problem and proposed two parameter selection methods using \textit{a priori} image sparsity estimates obtained from the k-space data for temporal sparsity and from a reference image for spatial sparsity. The evaluations of the methods were carried out using both simulated and experimental golden angle DCE-MRI data. The two parameter selection methods proposed were the S-surface and S-curve methods. The proposed methods were compared with the L-curve and MC-SURE parameter selection methods. In the simulated test case, the parameter selection methods were also compared with a parameter pair that gives the smallest joint RMS reconstruction error with respect to the true target images.
	
	Both proposed parameter estimation methods produced a parameter pair that is close to the parameter pair with the minimum joint RMSE in the simulated test cases, and the RMS reconstruction errors with these parameter selections were only slightly larger than the minimum RMSE, especially with the S-curve method. With 10\% noise, the S-surface and S-curve methods both yielded a bit too high spatial regularization parameter leading to a larger than optimal joint RMS error. In the simulated test cases the L-curve yielded the worst parameter selections with all the different noise levels. The MC-SURE parameter selection performance in the joint RMSE measure in the simulated test case was slightly worse than with the S-curve, but slightly better than the S-surface.
	
	In the experimental case there is no minimum RMSE solution or other known optimal reference solution available, but the parameter pairs obtained with the proposed methods are close to a manually selected parameter pair. All the reconstructions are visually similar with the L-curve and MC-SURE reconstructions being slightly more noisy, and having more signal fluctuation.
	
	While the S-surface method is based on a 2D search which requires reconstructions over a grid of $P\times L$ parameter pairs $(\alpha,\beta)$, the S-curve method is based on two 1D searches, and therefore it needs only $P+L+1$ reconstructions making it computationally more efficient than the S-surface method. The S-curve method is also computationally more efficient than the MC-SURE method, which requires the computation of two reconstructions for each parameter pair, i.e. requiring $2(P+L)$ reconstructions in total. This suggests that the S-curve based selection could be a more favourable choice, especially in large scale 4D problems, as the methods produced similar parameter selection and reconstruction accuracy.
	
	As the expected temporal sparsity level $\hattvt$ for the selection of the temporal regularization parameter is estimated from differences of the zero frequency components of the measured k-space data between consecutive time steps, the estimate provides an approximation for the temporal total variation of the unknown image series. In the simulated test case with 5\% noise, the discrepancy between the true temporal TV of the ground truth images and the estimate $\hattvt$ obtained from the noisy k-space data using \eqref{tvtest} was approximately $22 \%$ due to the noise level in the simulated k-space data. Despite this discrepancy, the proposed parameter selections yielded highly feasible regularization parameters. However, we remark that in cases where the images between consecutive time steps exhibit changes due to motion or large opposite signed contrast changes that would lead to only small changes in the zero frequency coefficients, the data based sparsity estimate could be clearly smaller than the actual temporal TV, potentially leading to a less optimal parameter choice.

	%%%%%%%%%%%%%%%%%%%%%%%%%%%%%%%%%%%%%%%%%%%%%%%%%%%%%%%%%%%%%%%%%
	\section{Conclusions.}
	
	In this work, we proposed to use two sparsity based methods based on \textit{a priori} sparsity estimates for the automatic selection of the regularization parameters for the TV regularized CS problem. The S-surface method selects the regularization parameters based on the expected sparsity of unknown images in the two domains of regularization simultaneously. We also adopted a faster method called the S-curve, where the regularization parameters were selected one at a time, which was shown to produce similar reconstruction accuracy and parameter selection. The S-curve method requires the computation of fewer reconstructions making it computationally more efficient.
	
	Both of the approaches were demonstrated to lead to a highly feasible choice of the temporal and spatial regularization parameters in both the simulated and the experimental DCE-MRI experiment of a rat brain.
	
	\section*{Acknowledgments} 
	This work was supported by the Academy of Finland (Projects 312343 and 336791, Finnish Centre of Excellence in Inverse Modelling and Imaging, 2018--2025), the Jane and Aatos Erkko Foundation and the Väisälä Fund.
	
	\printbibliography

\end{document}